\documentclass[amsmath,amssymb,aps,pra,reprint,groupedaddress,showpacs,superscriptaddress]{revtex4-1}

\usepackage{multirow}
\usepackage{verbatim}

\newcommand{\Proof}{\noindent\textbf{Proof.}\quad}
\newcommand{\qed}{\hfill$\Box$}
\newcommand{\ket}[1]{\left\vert #1 \right\rangle}

\newtheorem{theorem}{Theorem}
\newtheorem{lemma}[theorem]{Lemma}

\newtheorem{proposition}[theorem]{Proposition}

\begin{document}

\title{Algebraic techniques in designing quantum synchronizable codes}

\author{Yuichiro Fujiwara}
\email[]{yuichiro.fujiwara@caltech.edu}
\affiliation{Division of Physics, Mathematics and Astronomy, California Institute of Technology, MC 253-37, Pasadena, California 91125, USA}

\author{Vladimir D. Tonchev}
\affiliation{Department of Mathematical Sciences, Michigan Technological University, Houghton, MI 49931, USA}

\author{Tony W. H. Wong}
\affiliation{Division of Physics, Mathematics and Astronomy, California Institute of Technology, MC 253-37, Pasadena, California 91125, USA}

\date{\today}

\begin{abstract}
Quantum synchronizable codes are quantum error-correcting codes
that can correct the effects of quantum noise as well as block synchronization errors.
We improve the previously known general framework for designing quantum synchronizable codes through more extensive use of the theory of finite fields.
This makes it possible to widen the range of tolerable magnitude of block synchronization errors
while giving mathematical insight into the algebraic mechanism of synchronization recovery.
Also given are families of quantum synchronizable codes based on punctured Reed--Muller codes and their ambient spaces.
\end{abstract}


\maketitle

\section{Introduction}\label{intro}
Quantum error correction is a fundamental tool in quantum information science
that allows for quantum information processing in a noisy environment.
Quantum noise is typically described by operators that act on qubits,
with the most general model being the linear combinations of the Pauli operators $I$, $X$, $Y$, and $Z$ acting on each qubit \cite{KLV}.
In this sense, quantum error-correcting codes can be seen as coding techniques that allow for recovering the original quantum state
when unintended operators may act on some qubits.

Active quantum error detection is an important method for suppressing quantum noise,
where one extracts the information about what kind of quantum error occurred on which qubit
through measurement without learning anything about the quantum information carried by qubits.
With this information, the effect of quantum noise can be reversed by applying appropriate quantum operations.

Very recently, a scheme that actively deals with a different type of error
due to misalignment with respect to the block structure of a qubit stream was introduced \cite{fblock}.
To describe the kind of misalignment the scheme considers,
assume that we have three qubits $q_0$, $q_1$, $q_2$
and encode each of them by the perfect $5$-qubit code given in \cite{MikeandIke} (see \cite{LMPZ,BDSW} for different realizations of the perfect $5$-qubit code).
Then the quantum information we have can be expressed by a sequence of fifteen qubits,
where each $5$-qubit state $\ket{\psi_i}$, $i=0, 1, 2$, represents one logical qubit of quantum information that corresponds to the original qubit $q_i$.
In order to correctly process quantum information,
we need to know the exact location of the boundary of each $5$-qubit block in the $15$-qubit state $\ket{\psi_0}\ket{\psi_1}\ket{\psi_2}$.
For instance, if misalignment occurs by two qubits to the left when handling the stream of fifteen qubits,
a quantum device trying to correct quantum errors on $\ket{\psi_1}$
will apply the quantum operation on the wrong set of five qubits, two of which come from $\ket{\psi_0}$ and three of which belong to $\ket{\psi_1}$.
More complicated examples involving other types of errors include failure in detecting photons at the beginning of photonic quantum communication, where
the receiver misses the first photon at the start of communications
and wrongly assumes that the following five photons that are properly detected form an encoded $5$-qubit block.
In this case, misalignment and a quantum error due to qubit loss occur simultaneously.

The current paper studies a coding scheme that allows for extracting the information about
the magnitude and direction of misalignment through non-disturbing measurement
while simultaneously figuring out the types and positions of standard quantum errors on qubits.
In other words, we investigate a quantum analogue of \textit{synchronizable error-correcting codes} \cite{BC}.

More formally, a coding scheme is called a \textit{quantum synchronizable} $(a_l, a_r)$-$[[n,k]]$ \textit{code}
if it encodes $k$ logical qubits into $n$ physical qubits and
corrects misalignment by up to $a_l$ qubits to the left and up to $a_r$ qubits to the right.
To seamlessly achieve quantum error correction and synchronization recovery,
we would like quantum synchronizable codes to correct linear combinations of $I$, $X$, $Z$, and $Y$ that act on physical qubits as well.
For this task, the known quantum synchronizable error-correcting scheme
employs essentially the same two-step quantum error correction procedure as that for Calderbank-Shor-Steane (CSS) codes \cite{CS,Steane}.
Hence, in addition to misalignment, the scheme handles discretized bit and phase errors in two separate steps.

The known general method for constructing quantum synchronizable error-correcting codes
directly exploits special classical codes over the finite field $\mathbb{F}_2$ of order two.
A \textit{binary linear} $[n,k,d]$ {code} of \textit{length} $n$, \textit{dimension} $k$, and \textit{minimum distance} $d$
is a $k$-dimensional subspace $\mathcal{L}$ of the $n$-dimensional vector space $\mathbb{F}_2^n$
such that $\min\{\text{wt}(\boldsymbol{v}) \ \vert\ \boldsymbol{v} \in \mathcal{L}, \boldsymbol{v} \not= \boldsymbol{0}\} = d$,
where $\text{wt}(\boldsymbol{v})$ is the number of nonzero coordinates of $\boldsymbol{v}$.
In what follows, we always assume that classical codes are over $\mathbb{F}_2$ and omit the term binary.
A \textit{cyclic} $[n,k,d]$ code $\mathcal{C}$ is a linear $[n,k,d]$ code with the property that
every cyclic shift of every codeword $\boldsymbol{c}  = (c_0,\dots, c_{n-1}) \in \mathcal{C}$ is also a codeword.
Let $\mathcal{C}$ and $\mathcal{D}$ be two linear codes of the same length.
$\mathcal{D}$ is $\mathcal{C}$-\textit{containing} if $\mathcal{C} \subseteq \mathcal{D}$.
It is \textit{dual-containing} if it contains its dual $\mathcal{D}^{\perp} = \{\boldsymbol{d}^{\perp} \in \mathbb{F}_2^n \ \vert\ 
\boldsymbol{d} \cdot \boldsymbol{d}^{\perp} = \boldsymbol{0}\ \text{for all}\ \boldsymbol{d} \in \mathcal{D}\}$.

The known general framework for constructing quantum synchronizable codes rely on cyclic codes with special containing properties:
\begin{theorem}[\cite{fblock}]\label{generalconst}
If there exist a dual-containing cyclic $[n, k_1,d_1]$ code $\mathcal{C}$ and a $\mathcal{C}$-containing cyclic $[n, k_2, d_2]$ code with $k_1 < k_2$,
then for any pair $a_l$, $a_r$ of nonnegative integers satisfying $a_l + a_r < k_2 - k_1$
there exists a quantum synchronizable $(a_l, a_r)$-$[[n+a_l+a_r, 2k_1-n]]$ code that corrects
at least up to $\lfloor\frac{d_1-1}{2}\rfloor$ phase errors and
at least up to $\lfloor\frac{d_2-1}{2}\rfloor$ bit errors.
\end{theorem}
Note that if a linear code $\mathcal{C}$ is dual-containing, a $\mathcal{C}$-containing linear code is also dual-containing
\footnote{Assume that we have a pair $\mathcal{C}$, $\mathcal{D}$ of linear codes such that
$\mathcal{C}^{\perp} \subseteq \mathcal{C} \subset \mathcal{D}$.
Then, because $\mathcal{C} \subset \mathcal{D}$ if and only if  $\mathcal{D}^{\perp} \subset \mathcal{C}^{\perp}$,
we have $\mathcal{D}^{\perp} \subset \mathcal{C}^{\perp} \subseteq \mathcal{C} \subset \mathcal{D}$.
Hence, we have $\mathcal{D}^{\perp} \subset \mathcal{D}$.}.
Hence, what the above theorem requires is actually a pair of dual-containing cyclic codes, one of which is strictly contained in another
and both of which guarantee large minimum distances.
While it is already a challenging problem to construct cyclic codes with good minimum distances,
it is not impossible to find infinitely many nontrivial examples satisfying the additional stringent conditions.
The following is the family of quantum synchronizable codes explicitly mentioned in the literature.
\begin{theorem}[\cite{fblock}]\label{corobch}
Let $n$, $d_1$, and $d_2$ be odd integers satisfying $n=2^m -1$ and $3 \leq d_2 < d_1 \leq 2^{\lceil \frac{m}{2} \rceil}-1$,
where $m \geq 5$.
Then for some $d_1' \geq d_1$, some $d_2' \geq d_2$, and any pair $a_l$, $a_r$ of nonnegative integers satisfying $a_l+a_r < \frac{m(d_1-d_2)}{2}$
there exists a quantum synchronizable $(a_l, a_r)$-$[[n+a_l+a_r,n-m(d_2-1)]]$ code that corrects
at least up to $\frac{d_1'-1}{2}$ phase errors and
at least up to $\frac{d_2'-1}{2}$ bit errors.
\end{theorem}

The primary purpose of the present paper is to improve the code design framework given in Theorem \ref{generalconst}
through careful analysis of the algebraic machinery behind synchronization recovery
as well as to give families of quantum synchronizable codes that are different from the one given in Theorem \ref{corobch}.
Our refined framework naturally improves the synchronization recovery capabilities
achievable by quantum synchronizable codes even if we use the same cyclic codes as the ones employed in Theorem \ref{corobch}.

In the next section, we briefly review quantum synchronizable coding that forms the basis of Theorem \ref{generalconst}
and give a precise description of one key aspect in the form of a mathematical lemma.
The coding scheme is reanalyzed in Section \ref{improvement} to improve its synchronization recovery capabilities.
Section \ref{pRM} enriches realizable parameters by giving families of quantum synchronizable codes
based on cyclic codes that have not previously been employed in the context of synchronization recovery.
Concluding remarks are given in Section \ref{conclusion}.

\section{Overview of block synchronization for qubits}\label{review}
Here we review the basics of block synchronization recovery for quantum information.
The simple mathematical model considered in \cite{fblock} is explained in Section \ref{model}.
Section \ref{mechanism} provides the overview and necessary mathematical details of quantum synchronizable coding.

\subsection{Preliminaries}\label{model}
Let $Q = (q_0,\dots, q_{x-1})$ be an ordered set of length $x$, where each element represents a qubit.
A \textit{block} $F_i$ is a set of consecutive elements of $Q$.
Let $\mathcal{F} = \{F_0, \dots, F_{y-1}\}$ be a set of blocks.
The ordered set $(Q, \mathcal{F})$ is called a \textit{block-wise structured sequence} if $|\bigcup_i F_i| = x$ and $F_i \cap F_j = \emptyset$ for $i \not= j$.
In other words, the elements of a sequence are partitioned into groups of consecutive elements called blocks.

Take a set $G = \{q_j, \dots, q_{j+g-1}\}$ of $g$ consecutive elements of $Q$.
The set $G$ is said to be \textit{misaligned} by $a$ qubits to the \textit{right} with respect to $(Q, \mathcal{F})$
if there exits an integer $a$ and a block $F_i$ such that $F_i = \{q_{j-a}, \dots, q_{j+g-a-1}\}$ and $G \not\in \mathcal{F}$.
If $a$ is negative, we may say that $G$ is misaligned by $\vert a \vert$ qubits to the \textit{left}.
$G$ is \textit{properly aligned} if $G \in \mathcal{F}$.

With this simple model, the three $5$-qubit blocks given as an example in the previous section 
may be seen as $Q = (q'_0, \dots, q'_{14})$, where the three encoded $5$-qubit blocks
$\ket{\psi_0}$, $\ket{\psi_1}$, and $\ket{\psi_2}$ form blocks
$F_0 = (q'_0, \dots, q'_4)$, $F_1 = (q'_5, \dots, q'_{9})$, and $F_2 = (q'_{10}, \dots, q'_{14})$ respectively.
These $15$ qubits are subject to quantum information processing
and may be sent to a different place, stored in quantum memory or immediately processed for quantum computation.

If misalignment occurs by, for instance, two qubits to the left during quantum error correction on $\ket{\psi_1}$,
the device applies the quantum error correction procedure to the set $G$ of five qubits $q'_3$, \dots, $q'_{7}$,
two of which come from $F_0$ and three of which belong to $F_1$.
For example, when measuring the stabilizer generator $XZZXI$ of the $5$-qubit code to obtain the syndrome,
the operation the device actually performs to the whole system can be expressed as
\[I^{\otimes 3}XZZXI^{\otimes 8}\ket{\psi_0}\ket{\psi_1}\ket{\psi_2},\]
which, if block synchronization were correct, would be
\[I^{\otimes 5}XZZXI^{\otimes 6}\ket{\psi_0}\ket{\psi_1}\ket{\psi_2}.\]
The operator $I^{\otimes 3}XZ$ does not stabilize $\ket{\psi_0}$, nor does $ZXI^{\otimes 3}$ $\ket{\psi_1}$.
Thus, the measurement process not only fails to obtain the correct syndrome but also introduces errors to the system.
Similarly, if the same misalignment happens during fault-tolerant quantum computation,
the device trying to perform the logical $\bar{X}$ operation applies $I^{\otimes 3}XX$ on the first $5$-qubit block
and $XXXI^{\otimes 2}$ on the next $5$-qubit block.

The goal of quantum synchronizable coding is to make it possible to extract the information about
how many qubits away the window is from the proper alignment and in which direction should misalignment occur
while keeping the quantum information carried by qubits intact.
For the sake of simplicity, we assume that a device regains access to all the qubits in proper order in the system
if misalignment is correctly detected and identified.

\subsection{Quantum synchronizable coding}\label{mechanism}
In this subsection we briefly review the mechanism of quantum synchronizable codes introduced in \cite{fblock}
and prove a lemma, which we will use in Section \ref{improvement}.
We assume familiarity with the structure of CSS codes and their encoding and decoding methods.
For the basic facts and notions in classical and quantum coding theories, the reader is referred to \cite{MS,MikeandIke}.

As defined in Section \ref{intro}, a cyclic code $\mathcal{C}$ of length $n$ is a linear code with the property that
if $\boldsymbol{c} = (c_0,\dots, c_{n-1})$ is a codeword of $\mathcal{C}$, then so is the cyclic shift $(c_{n-1},c_{0},\dots, c_{n-2})$.
It is known that, by regarding each codeword as the coefficient vector of a polynomial in $\mathbb{F}_2[x]$,
a cyclic code of length $n$ can be seen as a principal ideal in the ring $\mathbb{F}_2[x]/(x^n-1)$
generated by the unique monic nonzero polynomial $g(x)$ of minimum degree in the code which divides $x^n-1$.
When a cyclic code is of length $n$ and dimension $k$,
the set of codewords can be written as $\mathcal{C} = \{i(x)g(x) \ \vert \ \deg(i(x)) < k\}$, where the degree $\deg(g(x))$ of the generator polynomial is $n-k$.
A cyclic shift of a codeword naturally corresponds to multiplying by $x$ modulo $x^n-1$, which is an automorphism of the code.
The orbit of a given codeword $i(x)g(x)$ by this group action is written as $Orb_x(i(x)g(x)) = \{x^ai(x)g(x) \pmod{x^n-1} \ \vert \ a \in \mathbb{N}\}$,
where $\mathbb{N}$ is the set of positive integers.

Let $\mathcal{C}$ be a linear $[n,k_1,d_1]$ code.
Recall that a linear $[n,k_2,d_2]$ code $\mathcal{D}$ is said to be $\mathcal{C}^{\perp}$-containing if $\mathcal{C}^{\perp} \subseteq \mathcal{D}$.
The CSS construction turns a $\mathcal{C}^{\perp}$-containing linear code $\mathcal{D}$ into a quantum error-correcting $[[n,k_2-k_1]]$ code
capable of correcting up to $d_1$ phase errors and up to $d_2$ bit errors through the standard two-step decoding procedure.
The framework on which Theorem \ref{generalconst} is built exploits this quantum error correction mechanism,
as is suggested by the fact that the theorem requires a pair of cyclic codes $\mathcal{C}$ and $\mathcal{D}$
satisfying $\mathcal{C}^{\perp} \subseteq \mathcal{C} \subset \mathcal{D}$.

Let $\mathcal{C}$ be a dual-containing cyclic $[n, k_1,d_1]$ code contained in another cyclic $[n, k_2, d_2]$ code $\mathcal{D}$ with $k_1 < k_2$.
Define $g(x)$ as the the generator polynomial of $\mathcal{D}$
which is the unique monic nonzero polynomial of minimum degree in $\mathcal{D}$.
Define also $h(x)$ as the generator polynomial of $\mathcal{C}$ which is the unique monic nonzero polynomial of minimum degree in $\mathcal{C}$.
Since $\mathcal{C} \subset \mathcal{D}$,
the generator polynomial $g(x)$ divides every codeword of $\mathcal{C}$,
which means that $h(x)$ can be written as $h(x) = f(x)g(x)$ for some polynomial $f(x)$ of degree $n-k_1-\deg(g(x)) = k_2-k_1$.

For a polynomial $j(x) = j_0 + j_1x +\cdots+j_{n-1}x^{n-1}$ of degree less than $n$ over $\mathbb{F}_2$,
define $\ket{j(x)}$ as the $n$-qubit quantum state
$\ket{j(x)} = \ket{j_0}\ket{j_1}\cdots\ket{j_{n-1}}$.
For a set $J$ of polynomials of degree less than $n$ over $\mathbb{F}_2$, we define $\ket{J}$ as
\[\ket{J} = \frac{1}{\left\vert J \right\vert}\sum_{j(x) \in J}\ket{j(x)}.\]
Addition between $J$ and polynomial $k(x) \in \mathbb{F}_2[x]$ is defined as $J + k(x) = \{j(x)+k(x) \ \vert \ j(x) \in J\}$.

Let $R = \{r_i(x)\ \vert \ 0 \leq i \leq 2^{2k_1-n-1}\}$ be a system of representatives of the cosets $\mathcal{C}/\mathcal{C}^{\perp}$.
Take the set $V_g = \left\{\ket{\mathcal{C}^{\perp} + r_i(x) + g(x)} \ \vert \ r_i(x) \in R\right\}$ of $2^{2k_1-n}$ states.
Because $R$ is a system of representatives,
these $2^{2k_1-n}$ states form an orthonormal basis.
Let $\mathcal{V}_g$ be the vector space of dimension $2^{2k_1-n}$ spanned by $V_g$.
This space $\mathcal{V}_g$ plays the key role in extracting the information about the magnitude and direction of a synchronization error
through non-disturbing measurement.

\subsubsection{Encoding}\label{enc}
Take a parity-check matrix $H_{\mathcal{D}}$ of $\mathcal{D}$.
We assume that $H_{\mathcal{D}}$ is of full rank.
For each row of $H_{\mathcal{D}}$, replace zeros with $I$s and ones with $X$s.
Perform the same replacement with $I$s for zeros and $Z$s for ones.
Because the condition that $\mathcal{C}^{\perp}\subseteq\mathcal{C}\subset\mathcal{D}$ implies $\mathcal{D}^{\perp}\subset\mathcal{D}$,
the code $\mathcal{D}$ is a dual-containing cyclic code of dimension $k_2$.
Hence, the resulting $2(n-k_2)$ Pauli operators on $n$ qubits form stabilizer generators $\mathcal{S}_{\mathcal{D}}$ of the Pauli group on $n$ qubits
that fixes a subspace of dimension $2^{k_2}$.
The set of the Pauli operators on $n$ qubits in $\mathcal{S}_{\mathcal{D}}$ that consist of $Z$s and $I$s is referred to as $\mathcal{S}_{\mathcal{D}}^Z$.
Construct stabilizer generators $\mathcal{S}_{\mathcal{C}}$ in the same way by using $\mathcal{C}$.

Take an arbitrary $(2k_1-n)$-qubit state $\ket{\varphi}$.
By using an encoder for the CSS code of parameters $[[n,2k_1-n]]$ defined by $\mathcal{S}_{\mathcal{C}}$,
we encode the state $\ket{\varphi}$ into $n$-qubit state
$\ket{\varphi}_{\text{enc}} = \sum_i\alpha_i\ket{\boldsymbol{v}_i}$,
where each $\boldsymbol{v}_i$ is an $n$-dimensional vector
with the orthogonal basis being $\left\{\ket{\mathcal{C}^{\perp} + r_i(x)} \ \vert \ r_i(x) \in R\right\}$.
Let $U_{\boldsymbol{g}}$ be the unitary operator that adds the coefficient vector $\boldsymbol{g}$ of the generator polynomial $g(x)$.
By applying $U_{\boldsymbol{g}}$, we have
$U_{\boldsymbol{g}} \ket{\varphi}_{\text{enc}} = \sum_i\alpha_i\ket{\boldsymbol{v}_i +\boldsymbol{g}}$.

To describe the final step of encoding, we need a notion from algebra.
Let $f(x) \in \mathbb{F}_2[x]$ be a polynomial over $\mathbb{F}_2$ such that $f(0)=1$.
The cardinality $\operatorname{ord}(f(x)) = \vert\{x^a \pmod{f(x)} \ \vert \ a \in \mathbb{N}\}\vert$ is called the \textit{order} of the polynomial $f(x)$.
This cardinality is also known as the \textit{period} or \textit{exponent} of $f(x)$.
Note that in our case the condition that $h(x)$ divides $x^n-1$ implies that its factor $f(x)$ also divides it,
which dictates that $\operatorname{ord}(f(x)) \leq n$.
In what follows, when we consider a representative of the equivalence class $f_0(x) \pmod{f_1(x)}$ for given two polynomials $f_0(x)$ and $f_1(x)$,
we choose the one with the smallest nonnegative degree, that is, the remainder of $f_0(x)$ divided by $f_1(x)$.

Take a pair $a_l$, $a_r$ of nonnegative integers such that $a_l+a_r < \operatorname{ord}(f(x))$.
Using $a_l+a_r$ ancilla qubits and CNOT gates, we take this state to an $(n+a_l+a_r)$-qubit state as follows:
\[\ket{0}^{\otimes a_l}U_{\boldsymbol{g}}\ket{\varphi}_{\text{enc}}\ket{0}^{\otimes a_r}
\rightarrow \sum_i\alpha_i\ket{\boldsymbol{w}_i^1, \boldsymbol{v}_i+\boldsymbol{g}, \boldsymbol{w}_i^2},\]
where $\boldsymbol{w}_i^1$ and $\boldsymbol{w}_i^2$ are the last $a_l$
and the first $a_r$ bits of the vector $\boldsymbol{v}_i +\boldsymbol{g}$ respectively.
The resulting encoded state
$\ket{\psi}_{\text{enc}} = \sum_i\alpha_i\ket{\boldsymbol{w}_i^1, \boldsymbol{v}_i+\boldsymbol{g}, \boldsymbol{w}_i^2}$
then goes through a noisy quantum channel.

\subsubsection{Decoding}\label{D}
To recover the original state $\ket{\varphi}$,
gather $n+a_l+a_r$ consecutive qubits $G = (q_0,\dots, q_{n+a_l+a_r-1} )$.
If block synchronization is correct, then $G$ is exactly the qubits of $\ket{\psi}_{\text{enc}}$ on which quantum errors may have occurred.
We assume the situation where $G$ can be misaligned by $a$ qubits to the right, where $-a_l \leq a \leq a_r$.

Let $P = (p_0,\dots, p_{n+a_l+a_r-1})$ be the $n+a_l+a_r$ qubits of the encoded state $\ket{\psi}_{\text{enc}}$.
Trivially, if $a=0$, then $P = G$.
Define $G_m = (q_{a_l}, \dots, q_{a_l + n -1})$.
By assumption, we have $G_m = (p_{a_l + a}, \dots, p_{a_l + n -1 + a})$.
Let $n$-fold tensor product $E$ of linear combinations of the Pauli matrices
be the errors that occurred on $P$.

We first correct bit errors that occurred on qubits in $G_m$ in the same manner as the separate two-step error correction procedure for a CSS code.
Because $\mathcal{C} \subset \mathcal{D}$,
the vector space spanned by the orthogonal basis stabilized by $\mathcal{S}_{\mathcal{D}}$ contains $\mathcal{V}_g$ as a subspace.
Hence, through a unitary transformation using $\mathcal{S}_{\mathcal{D}}^Z$, we can obtain the error syndrome for the window in the same way
as when detecting errors with the CSS code defined by $\mathcal{S}_{\mathcal{D}}$ as follows:
\[E\ket{\psi}_{\text{enc}}\ket{0}^{\otimes n-k_2}\\
\rightarrow E\ket{\psi}_{\text{enc}}\ket{\chi},\]
where $\ket{\chi}$ is the $(n-k_2)$-qubit syndrome by $\mathcal{S}_{\mathcal{D}}^Z$ (see \cite{fblock} for a rigorous proof).
If $E$ introduced at most $\lfloor \frac{d_2-1}{2} \rfloor$ bit errors on qubits in $G_m$,
these quantum errors are detected and then corrected by applying the $X$ operators accordingly.

Synchronization recovery is performed by taking advantage of the window $G_m$ on which all bit errors are corrected.
We describe the procedure as a proof of a lemma
that will play an important role in improving the maximum tolerable magnitude of synchronization errors.
\begin{lemma}\label{uniqueness}
Let $\mathcal{C}$ be a dual-containing cyclic code of length $n$ and dimension $k_1$
and $\mathcal{D}$ a $\mathcal{C}$-containing cyclic code of the same length, larger dimension $k_2 > k_1$, and minimum distance $d_2$.
Assume that $h(x)$ and $g(x)$ are the generator polynomials of $\mathcal{C}$ and $\mathcal{D}$ respectively.
Define polynomial $f(x)$ of degree $k_2-k_1$ to be the factor of $h(x)$ such that $h(x) = f(x)g(x)$ over $\mathbb{F}_2[x]/(x^n-1)$.
Then for every pair $a_l$, $a_r$ of nonnegative integers such that $a_l+a_r < \operatorname{ord}(f(x))$
there exists a quantum synchronizable $(a_l, a_r)$-$[[n+a_l+a_r, 2k_1-n]]$ code
under the assumption that no sequence of consecutive $n$ qubits suffers from more than $\left\lfloor\frac{d_2-1}{2}\right\rfloor$ bit errors.
\end{lemma}
\Proof
Encode an arbitrary $(2k_1-n)$-qubit state $\ket{\varphi}$ by using a pair $\mathcal{C}$, $\mathcal{D}$ of cyclic codes
such that $\mathcal{C}^{\perp} \subseteq \mathcal{C} \subset \mathcal{D}$ as described in Section \ref{enc}.
Let operator $E$ be the quantum noise introduced to the encoded state $\ket{\psi}_{\text{enc}}$.
We assume the situation where misalignment occurred by $a$ qubits to the right
with the condition that $-a_l \leq a \leq a_r$, where the two nonnegative integers satisfy the inequality $a_l+a_r < \operatorname{ord}(f(x))$.
Perform the bit error correction on window $G_m$ as described earlier in Section \ref{D}.
These transformations can be expressed as
\begin{align*}
\ket{\varphi} &\rightarrow \ket{\psi}_{\text{enc}}\\
&\rightarrow E\ket{\psi}_{\text{enc}}\\
&\rightarrow E'\ket{\psi}_{\text{enc}},
\end{align*}
where operator $E'$ represents the partially corrected quantum errors after bit error correction on $G_m$.
Recall that all codewords of $\mathcal{C}^{\perp}$ and $r_i(x) \in R$ are also codewords of $\mathcal{C}$, and hence of $\mathcal{D}$ as well.
Because the polynomial $g(x)$ is the generator of $\mathcal{D}$,
it divides any polynomial of the form $s(x) + r_i(x) + g(x)$ over $\mathbb{F}_2[x]/(x^n-1)$, where $s(x) \in \mathcal{C}^{\perp}$.
Since we have
\[s(x) + r_i(x) + g(x) = i_0(x)f(x)g(x)+i_1(x)f(x)g(x)+g(x)\]
for some polynomials $i_0(x)$ and $i_1(x)$ whose degrees are both less than $k_1$,
the quotient is of the form $j(x)f(x)+1$ for some polynomial $j(x)$.
Dividing the quotient by $f(x)$ gives $1$ as the remainder.
It is easy to show that $\left\vert Orb_x(g(x)) \right\vert = n$ (see \cite{fblock} for an elementary proof).
Thus, applying the same two-step division procedure to any polynomial appearing as a state
in cyclically shifted $V_g$ by $a$ qubits gives the reminder of $x^a$ divided by $f(x)$ in $\mathbb{F}_2[x]/(x^n-1)$.
Because $h(x)$ divides $x^n-1$, its factor $f(x)$ also divides $x^n-1$.
Hence, the resulting remainder is exactly the representative of $x^a \pmod{f(x)}$ with a nonnegative degree less than $k_2-k_1$.
Note that every state in $V_g $ is of the form $\ket{\mathcal{C}^{\perp} + r_i(x) + g(x)}$.
If $G_m$ contains no bit errors after bit error correction,
the basis states of the corresponding portion in $E'\ket{\psi}_{\text{enc}}$ are the cyclically shifted coefficient vectors of the correct polynomials.
Let $Q_{t(x)}$ and $R_{t(x)}$ be polynomial division operations on $n$ qubits that give
the quotient and remainder respectively through quantum shift registers defined by a polynomial $t(x)$ of degree less than $n$ \cite{GB}.
Let $\mathfrak{Q} = I^{\otimes a_l + a}Q_{g(x)}I^{\otimes a_r - a}$
and $\mathfrak{R} = I^{\otimes n + a_l + a_r}R_{f(x)}$,
so that the two represent applying $Q_{g(x)}$ to the window and $R_{f(x)}$ to the ancilla qubits of $Q_{g(x)}$ that contain the calculated quotient.
This pair of operations give the syndrome for the synchronization error as
\[E'\ket{\psi}_{\text{enc}} \ket{0}^{\otimes n}
\xrightarrow{\mathfrak{R}\mathfrak{Q}}
E'\ket{\psi}_{\text{enc}} \ket{x^a\ (\text{mod}\ {f(x)})},\]
where $\left\vert 0 \right\rangle^{\otimes n}$ is the ancilla for $Q_{g(x)}$
and $\ket{x^a\ (\text{mod}\ {f(x)})}$ is the state defined by the representative of $x^a\ (\text{mod}\ {f(x)})$.
If $x^b \not\equiv x^{c} \pmod{f(x)}$ for any pair $b$, $c$ of distinct nonnegative integers less than or equal to $a_l+a_r$,
the remainder given as the representative of $x^a \pmod{f(x)}$ uniquely identifies the magnitude and direction of the synchronization error $a$.
By assumption, we have $a_l+a_r < \operatorname{ord}(f(x))$.
Thus, we have the cardinality
\[\vert\{x^a\ (\text{mod}{f(x)}) \ \vert\ 0 \leq a \leq a_l+a_r\}\vert = a_l+a_r+1\]
as desired. The proof is complete.
\qed

With the procedure described in the proof above, we can obtain the information about
how many qubits away $G = (q_0,\dots, q_{n+a_l+a_r-1} )$ is from the proper position $P = (p_0,\dots, p_{n+a_l+a_r-1})$ and in which direction.
Thus, by assumption, we can correctly shift the window to the last $n$ qubits $(p_{a_l+a_r},\dots,p_{n+a_l+a_r-1})$ of $P$.
Because we employed classical cyclic codes,
the same error correction procedure can be performed on $(p_{a_l+a_r},\dots,p_{n+a_l+a_r-1})$,
allowing for correcting bit errors that may have occurred on the last $n$ qubits of $P$.
By the same token, moving the window to the first $n$ qubits of $P$ allows us to correct the remaining bit errors on $P$.
Hence, if the channel introduced at most $\lfloor \frac{d_2-1}{2} \rfloor$ bit errors on any consecutive $n$ quibits,
we can correct all bit errors that occurred on the qubits in $P$.

The remaining decoding procedure for recovering the original $(2k_1-n)$-qubit state $\ket{\varphi}$
is to shrink the $(n+a_l+a_r)$-qubit state while correcting phase errors.
This can be done by running backwards the translation and expansion operations we applied to $\ket{\varphi}_{\text{enc}}$
and then applying a decoding circuit of the CSS code
based on the dual-containing cyclic code $\mathcal{C}$ (see \cite{fblock} for details).

\section{Improving synchronization error tolerance}\label{improvement}
In this section we examine the maximum tolerable magnitude of synchronization errors.

The reason that Theorem \ref{generalconst} can only tolerate up to a $(k_2 - k_1-1)$-qubit shift is that
the original proof given in \cite{fblock} does not use the concept of the order of a polynomial.
In fact, in view of Lemma \ref{uniqueness},
the original proof can be understood as a naive application of a rather conservative lower bound on the order of $f(x)$,
namely $\operatorname{ord}(f(x)) \geq \deg(f(x))$.
Here we aim to improve synchronization recovery capabilities by examining the exact value of $\operatorname{ord}(f(x))$.

To avoid being overly general, we focus on the most relevant case where the code length is a Mersenne number $n = 2^m-1$.
This is because the known quantum synchronizable codes and the ones we will introduce in the next section all have lengths of this form.

\begin{theorem}\label{maintheorem}
Let $m$, $n$ be positive integers such that $n=2^m-1$,
and $\mathcal{C}$, $\mathcal{D}$ a dual-containing cyclic $[n, k_1,d_1]$ code
with generator polynomial $h(x)$ and $\mathcal{C}$-containing cyclic $[n, k_2, d_2]$ code with generator polynomial $g(x)$ respectively.
Define polynomial $f(x)$ of degree $k_2 - k_1$ as the quotient of $h(x) = f(x)g(x)$ divided by $g(x)$
and write its factorization into irreducible polynomials as $f(x) = \prod_if_i(x)$.
For every pair $a_l$, $a_r$ of nonnegative integers such that $a_l + a_r < \operatorname{lcm}_i\{\operatorname{ord}(f_i(x))\}$
there exists a quantum synchronizable $(a_l, a_r)$-$[[n+a_l+a_r, 2k_1-n]]$ code that corrects
at least up to $\lfloor\frac{d_1-1}{2}\rfloor$ phase errors and
at least up to $\lfloor\frac{d_2-1}{2}\rfloor$ bit errors.
In particular, the maximum tolerable magnitude $\operatorname{lcm}_i\{\operatorname{ord}(f_i(x))\}-1$ attains $n-1$, which is the largest possible,
if $f(x)$ has a primitive polynomial $f_i(x)$ of degree $m$ as its factor.
\end{theorem}

To prove the above theorem, we employ the following four facts in finite fields.
\begin{proposition}\label{fact1}
Let $m$ be a positive integer and $f(x)$
the product of all irreducible polynomials over $\mathbb{F}_2$ whose degrees divide $m$.
Then 
\[f(x) = x^{2^m}-x.\]
\end{proposition}
\begin{proposition}\label{fact2}
Let $f(x) = \prod_if_i(x)$ be a polynomial over $\mathbb{F}_2$, where $f_i(x)$ are all nonzero and pairwise relatively prime in $\mathbb{F}_2[x]$.
Then
\[\operatorname{ord}(f(x)) = \operatorname{lcm}_i\{\operatorname{ord}(f_i(x))\}.\]
\end{proposition}
\begin{proposition}\label{fact3}
If $f(x) \in \mathbb{F}_2[x]$ is an irreducible polynomial over $\mathbb{F}_2$,
then $\operatorname{ord}(f(x))$ divides $2^{\deg(f(x))}-1$.
\end{proposition}
\begin{proposition}\label{fact4}
A polynomial $f(x) \in \mathbb{F}_2[x]$ is primitive if and only if $f(0) = 1$, and
\[\operatorname{ord}(f(x))=2^{\deg(f(x))}-1.\]
\end{proposition}
For the proofs of these propositions, we refer the reader to \cite[Theorems 3.20 and 3.9, Corollary 3.4, and Theorem 3.16]{finitefield}.\\

\noindent\textbf{Proof of Theorem \ref{maintheorem}.}\quad
By Lemma \ref{uniqueness} and the rest of the argument in Section \ref{mechanism},
we only need to prove that
$\operatorname{ord}(f(x)) = \operatorname{lcm}_i\{\operatorname{ord}(f_i(x))\}$
and that $\operatorname{lcm}_i\{\operatorname{ord}(f_i(x))\} = n$ if at least one irreducible factor $f_i(x)$ is primitive and of degree $m$.
As mentioned in the proof of Lemma \ref{uniqueness},
because $h(x)$ is the generator polynomial of a cyclic code of length $n$,
its factor $f(x)$ divides $x^n-1$.
Thus, by Proposition \ref{fact1}, all $f_i(x)$ in the factorization $f(x) = \prod_if_i(x)$ are distinct.
Hence, Proposition \ref{fact2} proves that the order of our $f(x)$ is indeed the least common multiple of the orders of its irreducible factors $f_i(x)$.
Assume that one of irreducible factors of $f(x)$ is primitive and of degree $m$.
Note that for a pair $a$, $b$ of positive integers, $2^a-1$ divides $2^b-1$ if and only if $a$ divides $b$.
Hence, by Proposition \ref{fact1} and the fact that $f(x)$ divides $x^n-1$,
for each $i$ the integer $2^{\deg(f_i(x))}-1$ divides $2^m-1= n$.
Because $f(x)$ divides $x^n-1$, we have $f_i(0) = 1$ for every $i$.
Thus, by Propositions \ref{fact3} and \ref{fact4}, we have
\begin{align*}
\operatorname{lcm}_i\{\operatorname{ord}(f_i(x))\} &= 2^m-1\\
&= n.
\end{align*}
This completes the proof.
\qed

Because $\operatorname{lcm}_i\{\operatorname{ord}(f_i(x))\}$ is always at least $k_2-k_1$,
Theorem \ref{maintheorem} provides better synchronization recovery capabilities than Theorem \ref{generalconst}.
For instance, when $n=2^m-1$ is a prime, $m$ must be a prime as well.
In this case, Proposition \ref{fact1} dictates that for each $i$ the degree $\deg(f_i(x))$ is either $1$ or $m$.
Because $x-1$ is the only irreducible polynomial of degree $1$ with a nonzero constant term,
if $\deg(f(x)) \geq 2$, we have $\operatorname{ord}(f(x)) = n$,
achieving the highest possible synchronization error tolerance.

\section{Quantum synchronizable codes from Reed-Muller codes}\label{pRM}
In this section we study two special classes of algebraic codes to give families of quantum synchronizable error-correcting codes.
The first class is a type of finite geometry code based on projective geometry
while the other class includes those used in Theorem \ref{corobch} as a subclass.
To make the connection to our quantum synchronizable scheme as clear as possible,
we define these classical codes by their generator polynomials with the minimum amount of mathematics.
The proofs of the basic facts we use can be found in \cite{MS}.
For more finite geometric and algebraic views of our cyclic codes,
the interested reader is referred to \cite{MS,AK2,Charpin}.

For a nonnegative integer $s$ and a positive integer $n$,
the \textit{cyclotomic coset} $C_{s,n}$ of $s$ modulo $n$ over $\mathbb{F}_2$ is the set
\[C_{s,n} = \{s2^i \mod{n} \ \vert \ i \in \mathbb{N}\}.\]
Since $C_{s,n} = C_{s',n}$ if $s' \in C_{s,n}$, we may take a system
\[S_n = \{\min\{t \ \vert \ t \in C_{s,n}\} \ \vert \ s \in \mathbb{N}\cup\{0\}\}\]
of representatives of the cyclotomic cosets by picking the smallest element from each set.
We call $S_n$ the \textit{canonical} system of representatives.
The integers modulo $n$ are partitioned into cyclotomic cosets as
\[\{0,1,\dots,n-1\} = \bigcup_{s \in S_n} C_{s,n}.\]
Let $\alpha$ be a primitive $n$th root of unity in $\mathbb{F}_{2^{\vert C_{1,n}\vert}}$.
The minimal polynomial $M_s(x)$ of $\alpha^s$ over $\mathbb{F}_2$ can be expressed as
\[M_s(x) = \prod_{i \in C_{s,n}}(x-\alpha^i).\]

For nonnegative integers $s$, let $\operatorname{w}_2(s)$ denote the number of $1$'s in the binary expansion of $s$.
For positive integers $r$, $m$ such that $r < m$,
the \textit{punctured Reed-Muller code} $\mathcal{R}(r,m)^*$ of \textit{order} $r$ over projective space $\textup{PG}(m-1,2)$ is
the cyclic code of parameters
\[\left[2^m-1, \sum_{i=0}^{r}{{m}\choose{i}}, 2^{m-r}-1\right]\]
defined by the generator polynomial
\[g(x) = \prod_{\substack{1 \leq \operatorname{w}_2(s) \leq m-r-1\\ s \in S_{2^m-1}}}M_s(x).\]
For a comprehensive treatment of punctured Reed-Muller codes, the interested reader is referred to \cite{MS}.
We use the basic property of $\mathcal{R}(r,m)^*$
that the generator polynomial $g^{\perp}(x)$ of its dual ${\mathcal{R}(r,m)^*}^{\perp}$ is
\[g^{\perp}(x) = (x+1)\prod_{\substack{1 \leq \operatorname{w}_2(s) \leq r\\ s \in S_{2^m-1}}}M_s(x).\]
Punctured Reed-Muller codes are cyclic codes with the desired nested property for our purpose:
\begin{lemma}\label{rmlemma}
For any positive integers $r_1$, $r_2$, and $m$ such that $\lceil \frac{m}{2} \rceil< r_2 < r_1 < m$,
the punctured Reed-Muller codes of order $r_1$ and $r_2$ over $\textup{PG}(m-1,2)$ satisfy the condition that
\[{\mathcal{R}(r_2,m)^*}^{\perp} \subseteq \mathcal{R}(r_2,m)^* \subset \mathcal{R}(r_1,m)^*.\]
\end{lemma}
\Proof
Let $g_1(x)$, $g_2(x)$, and $g_2^{\perp}(x)$ be the generator polynomials of
$\mathcal{R}(r_1,m)^*$, $\mathcal{R}(r_2,m)^*$, and ${\mathcal{R}(r_1,m)^*}^{\perp}$ respectively.
Because these are generators of the corresponding principal ideals of $\mathbb{F}_2[x]$,
we only need to show that $g_1(x)$ divides $g_2(x)$ and that $g_2(x)$ divides $g_2^{\perp}(x)$.
Because $r_2 < r_1$, we have
\[g_2(x)= g_1(x)\prod_{\substack{m-r_1 \leq \operatorname{w}_2(s) \leq m-r_2-1\\ s \in S_{2^m-1}}}M_s(x).\]
Because $\lceil \frac{m}{2} \rceil< r_2 < m$, we have
\[g_2^{\perp}(x) = g_2(x)(x+1)\prod_{\substack{m-r_2 \leq \operatorname{w}_2(s) \leq r_2 \\ s \in S_{2^m-1}}}M_s(x).\]
The proof is complete.
\qed

The above lemma allows us to use punctured Reed-Muller codes as the cyclic codes $\mathcal{C}$ and $\mathcal{D}$ in Theorem \ref{maintheorem}
to obtain a family of quantum synchronizable codes:

\begin{theorem}\label{rmtheorem}
Let $r_1$, $r_2$, $m$, and $n$ be positive integers such that $\lceil \frac{m}{2} \rceil< r_2 < r_1 < m$ and such that $n=2^m-1$.
For every pair $a_l$, $a_r$ of nonnegative integers such that $a_l + a_r < \operatorname{lcm}_s\{\operatorname{ord}(M_s(x))\}$,
where $s$ runs through all integers in the canonical system $S_n$ of representatives of cyclotomic cosets modulo $n$
satisfying the condition that $m-r_1 \leq \operatorname{w}_2(s) \leq m-r_2-1$,
there exists a quantum synchronizable $(a_l, a_r)$-$[[n+a_l+a_r, 2\sum_{i=0}^{r_2}{{m}\choose{i}}-n]]$ code that corrects
at least up to $2^{m-r_2-1}-1$ phase errors and
at least up to $2^{m-r_1-1}-1$ bit errors.
\end{theorem}

Another useful property of punctured Reed-Muller codes is that their ambient spaces contain well-known cyclic codes.
Let $n$ be an odd integer and $\alpha \in \mathbb{F}_{2^{\vert C_{1,n}\vert}}$ a primitive $n$th root of unity.
A \textit{Bose-Chaudhuri-Hocquenghem} (BCH) \textit{code} of \textit{length} $n$ and \textit{designed distance} $d$ is a cyclic code of length $n$
whose generator polynomial is
\[g(x) = \prod_{i \in \bigcup_{j=0}^{d-2}C_{b+j,n}}(x-\alpha^i),\]
where $b$ is a nonnegative integer.
The term designed distance reflects the fact that the true minimum distance of a BCH code is at least its designed distance.
The proof of this fact and other basic properties of BCH codes can be found in \cite{MS}.
A BCH code is \textit{primitive} if the length is of the form $n=2^m-1$ for some positive integer $m$,
and \textit{narrow-sense} if $b = 1$.

BCH codes are one of the older classes of cyclic codes and have extensively been studied in classical coding theory.
Their dual-containing property and basic parameters have also been investigated in the context of quantum error correction \cite{Steane2,AKS}.
For this reason, they have a great potential as a source of excellent quantum synchronizable codes.
In fact, Theorem \ref{corobch} is a straightforward application of primitive, narrow-sense BCH codes of odd designed distance.

We begin with the following observation.
\begin{lemma}\label{rmbch}
Let $\mathcal{B}$ be the primitive, narrow-sense \textup{BCH} code of length $2^m-1$ and designed distance $2^{m-r}-1$,
where $\lceil \frac{m}{2} \rceil< r < m-2$ and $m \geq 7$.
Then
\[{\mathcal{R}(r,m)^*}^{\perp} \subseteq \mathcal{R}(r,m)^* \subset \mathcal{B}.\]
\end{lemma}
\Proof
Let $S$ be the set of positive integers less than $2^{m-r}-1$.
Then the generator polynomial $g(x)$ of the primitive, narrow-sense BCH code of length $2^m-1$ and designed distance $2^{m-r}-1$ is
\begin{align*}
g(x) &= \prod_{i \in \bigcup_{j \in S}C_{j,2^m-1}}(x-\alpha^i)\\
&= \prod_{s \in S \cap S_{2^m-1}}M_s(x),
\end{align*}
where $ S_{2^m-1}$ is the canonical system of representatives of the cyclotomic cosets modulo $2^m-1$.
For any positive integer $a < 2^{m-r}-1$, we have $\operatorname{w}_2(a) \leq m-r-1$.
Hence, we have
\[S \cap S_{2^m-1} \subseteq \{s \ \vert \ s \in S_{2^m-1}, 1 \leq \operatorname{w}_2(s) \leq m-r-1\}.\]
Hence, $g(x)$ divides the generator polynomial of $\mathcal{R}(r,m)^*$,
which implies that $\mathcal{R}(r,m)^* \subseteq \mathcal{B}$.
It is known that $\mathcal{R}(r,m)^* \not= \mathcal{B}$ if $\lceil \frac{m}{2} \rceil< r < m-2$ and $m \geq 7$
(see, for example, \cite[Lemma 5.3]{Huffman}).
Hence, we have $\mathcal{R}(r,m)^* \subset \mathcal{B}$.
By Lemma \ref{rmlemma}, ${\mathcal{R}(r,m)^*}^{\perp} \subseteq \mathcal{R}(r,m)^*$.
The proof is complete.
\qed

Because BCH codes are cyclic, Lemma \ref{rmbch} states that we may use dual-containing punctured Reed-Muller codes together with
primitive, narrow-sense BCH codes to construct quantum synchronizable codes.
Note that $\mathcal{C}^{\perp} \subseteq \mathcal{C} \subset \mathcal{D}$ implies that $\mathcal{D}^{\perp} \subset \mathcal{D}$.
Since a BCH code is trivially contained in another BCH code of smaller designed distance,
we can also construct quantum synchronizable codes from a pair of dual-containing BCH codes without using punctured Reed-Muller codes.
The following are two useful known results on primitive, narrow-sense BCH codes that are dual-containing:
\begin{theorem}[\cite{AKS}]\label{dualcontainingbch}
For $m \geq 2$, a primitive, narrow-sense \textup{BCH} code of length $2^m-1$ is dual-containing if and only if
its designed distance $d$ satisfies the condition that $2 \leq d \leq 2^{\lceil \frac{m}{2} \rceil}-1$.
\end{theorem}
\begin{theorem}[\cite{AKS}]\label{bchdimension}
A primitive, narrow-sense \textup{BCH} code of length $2^m-1$ and designed distance $d$
that is dual-containing is of dimension $2^m-1-m\lceil\frac{d-1}{2}\rceil$.
\end{theorem}
By applying Lemmas \ref{rmlemma} and \ref{rmbch} and Theorems \ref{dualcontainingbch} and \ref{bchdimension} to Theorem \ref{maintheorem},
we can obtain a variety of quantum synchronizable codes.
For instance, the following is a special case based on punctured Reed-Muller codes and BCH codes:
\begin{theorem}\label{corormbch}
Let $n$, $r$, $m$ be positive integers satisfying the conditions that $n=2^m-1$ is a prime, that $\lceil \frac{m}{2} \rceil< r < m-2$, and that $m \geq 7$.
Then for any pair of nonnegative integers $a_l$, $a_r$ satisfying $a_l+a_r < n$
there exists a quantum synchronizable $(a_l, a_r)$-$[[n+a_l+a_r,\sum_{i=0}^{r}{{m}\choose{i}}]]$ code that corrects
at least up to $2^{m-r-1}-1$ phase errors and at least up to $2^{m-r-1}-1$ bit errors.
\end{theorem}

The following lemmas allow us to calculate the synchronization recovery capabilities
of quantum synchronizable error-correcting codes based on primitive, narrow-sense BCH codes:
\begin{lemma}[\cite{AKS}]\label{cosetsize}
Let $n$, $m$ be positive integers such that $n=2^m-1$.
For any positive integer $s \leq 2^{\lceil\frac{m}{2}\rceil}$,
the cardinality $\vert C_{s,n} \vert = m$.
\end{lemma}
\begin{lemma}[\cite{AKS}]\label{cosetdiff}
Let $n$, $m$ be positive integers such that $n=2^m-1$.
For any odd positive integer $s, s' \leq 2^{\lceil\frac{m}{2}\rceil}$,
we have $C_{s,n} \not= C_{s',n}$.
\end{lemma}

\begin{theorem}\label{theorem2bch}
Let $n$, $d_1$, and $d_2$ be odd integers satisfying $n=2^m -1$ and $3 \leq d_2 < d_1 \leq 2^{\lceil \frac{m}{2} \rceil}-1$,
where $m \geq 5$ and $d_1-d_2 \geq 4$.
Then for any pair of nonnegative integers $a_l$, $a_r$ satisfying $a_l+a_r < n$
there exists a quantum synchronizable $(a_l, a_r)$-$[[n+a_l+a_r,n-m(d_2-1)]]$ code that corrects
at least up to $\frac{d_1-1}{2}$ phase errors and at least up to $\frac{d_2-1}{2}$ bit errors.
\end{theorem}
\Proof
Apply Theorems \ref{dualcontainingbch} and \ref{bchdimension} to Theorem \ref{maintheorem}.
By Lemma \ref{cosetdiff} and the fact that for every positive even integer $s \leq d_1$ the cyclotomic coset $C_{s,n} = C_{\frac{s}{2},n}$,
the generator polynomials of the primitive, narrow-sense BCH codes of distance $d_1$ and $d_2$ are
\[g_1(x) = \prod_{\substack{1 \leq s \leq d_1-1 \\ s \ \text{odd}}}M_s(x)\]
and
\[g_2(x) = \prod_{\substack{1 \leq s \leq d_2-1 \\ s \ \text{odd}}}M_s(x)\]
respectively.
Thus, we only need to prove that
\[f(x) = \prod_{\substack{d_2 \leq s \leq d_1-1 \\ s \ \text{odd}}}M_s(x)\]
is of order $n$.
By Lemma \ref{cosetsize}, for $d_2 \leq s \leq d_1-1$ we have $\deg(M_s(x)) = m$.
Hence, because $\operatorname{ord}(M_s(x))$ is the order of $\alpha^s$ in the multiplicative group $\mathbb{F}_{2^m}^*$ (see \cite[Theorem 3.33]{finitefield}),
we have $\operatorname{ord}(M_s(x))=\frac{n}{\gcd(s,n)}$.
Because we have $d_1-d_2 \geq 4$, the polynomial $f(x)$ has two irreducible factors $M_s(x)$ and $M_{s+2}(x)$ for some odd $s$.
Hence, by Proposition \ref{fact2}, we have
\begin{align*}
\operatorname{ord}(f(x)) &\geq \operatorname{lcm}\left(\frac{n}{\gcd(s,n)},\frac{n}{\gcd(s+2,n)}\right)\\
&= n
\end{align*}
as desired. The proof is complete.
\qed

Since the parity of the designed distance of each BCH code in Theorem \ref{theorem2bch} does not affect whether
the pair of cyclic codes satisfy the nested property required to construct a quantum synchronizable code,
one may also exploit BCH codes of even designed distance to obtain similar quantum synchronizable error-correcting codes,
albeit of parameters slightly cumbersome to spell out.

\section{Concluding Remarks}\label{conclusion}

We refined the known general framework for designing quantum synchronizable codes through an algebraic approach.
With this refinement, we can compute the best attainable synchronization recovery capabilities a given pair of classical cyclic codes can offer.
We also examined the structures of punctured Reed-Muller codes and BCH codes in their ambient spaces
to obtain families of quantum synchronizable codes.

While we focused on the case when code lengths are of the form $n=2^m-1$,
in principle, we can also apply similar techniques to the general case when $n$ is a positive integer.
In fact, narrow-sense BCH codes that are not primitive are also known to be dual-containing
if their designed distances satisfy a condition similar to the one given in Theorem \ref{dualcontainingbch} \cite{AKS}.
The exact dimensions can be obtained in the same way as well.
Moreover, as we will see here,
our result on the maximum tolerable magnitude of misalignment can also be extended in theory to the case of general $n$.

To generalize our approach through Lemma \ref{uniqueness} to the case when $n$ may not be of the form $2^m-1$,
we need to know the order of a given polynomial $f(x)$
which divides $x^n-1$ but may contain irreducible factors of multiplicity more than one.
The following fact is useful for computing the order.
\begin{proposition}\label{multipleorder}
Let $f(x) \in \mathbb{F}_2[x]$ be irreducible over $\mathbb{F}_2$ with $f(0) = 1$ and $\operatorname{ord}(f(x))=e$.
Let $a$ be a positive integer and define $b$ to be the smallest integer such that $2^b \geq a$.
Then $\operatorname{ord}((f(x))^a) = 2^be$.
\end{proposition}
The proof of the above proposition can be found in \cite[Theorem 3.8]{finitefield}.

Because the polynomial of which we need to compute the order divides $x^n-1$,
its irreducible factors $f_i(x)$ all satisfy the condition that $f_i(0) = 1$.
Thus, by Propositions \ref{fact2} and \ref{multipleorder},
even if $n$ is not a Mersenne number,
the maximum tolerable magnitude of misalignment can be computed from the order of each irreducible factor.
A table of the orders of irreducible polynomials can be found in \cite{finitefield}.

Our block synchronization scheme may be seen as an algebraically modernized quantum analogue of the classical schemes introduced in the 60's,
where cosets of cyclic codes played the key role (see, for example, \cite{Levy,BC,TF}).
The theory of synchronization for classical bits has seen progress since its inception
and gave birth to different synchronization techniques.
The most recent major progress includes the proof of the existence of capacity achieving codes
in a single shot model within a finite length regime \cite{Polyanskiy}
and explicit constructions for high-rate self-synchronizing codes \cite{HighrateSC}.

A notable property of many newer classical codes for synchronization is that
they allow for locating boundaries regardless of the magnitude of misalignment while achieving high information rates.
This means that the sender and receiver can establish and maintain efficient communications over noisy channels
even if no prior block synchronization is assumed.
While such high level control over quantum information would be extremely challenging both theoretically and experimentally,
very recently an initial step from the theoretical side has been made in this direction as well \cite{parsing}.
It would be of interest to look for a way to realize quantum analogues of recent software solutions for synchronization in classical communications.

Another interesting related topic would be the type of synchronization error due not to misalignment but to undetected loss of bits.
Such synchronization errors are called \textit{deletions} in classical coding theory
(see \cite{MBT,Mitzenmacher} for surveys of results on this and closely related types of synchronization errors).
As far as the authors are aware, no result is available on the quantum analogue of this channel at the time of writing.
While a deletion can be recovered by our method in some cases such as qubit loss at the start of quantum communication,
quantum synchronizable codes are not able to treat all types of deletion.

In general, loss of qubits may be treated as \textit{erasures} or \textit{located errors} if there is a hardware solution for detecting such anomalies.
(see, for example, \cite{MB,LSH,NAC,LGZZYP,WB5,RHG,VWW,GKLVD}).
Hence, errors such as photon loss can be handled by tracing out the lost qubits and then recovering them
through quantum error-correcting codes for erasures such as those found in \cite{GBP}.
While this assumption is reasonable in many contexts such as linear optical quantum memories,
it may be more reasonable to also consider undetected loss of qubits in other contexts
such as asynchronous free-space optical quantum communication at high rates.
We hope that the present work will stimulate research on asynchronous quantum information transmission.

\begin{acknowledgments}
Y.F. acknowledges the support from JSPS. Vladimir Tonchev is supported by an NSA grant.
\end{acknowledgments}

\end{document}